\documentclass{iau} 
\usepackage{graphicx}

\def\etal       {et~al.}
\def\kms	{~km~s$^{-1}$}

\title[RRL Maser in MWC349A] 
{The Structure of the Radio Recombination Line Maser Emission in the Envelope of MWC349A}

\author[James M. Moran, Qizhou Zhang, \& Deanna L. Emery]   
{James M. Moran, Qizhou Zhang, and Deanna L. Emery}

\affiliation{Harvard-Smithsonian Center for Astrophysics, \\
Mail Stop 42, 60 Garden St., Cambridge, MA 02138, USA \\ email: {\tt jmoran@cfa.harvard.edu}}

\pubyear{2017}
\volume{336}
\jname{Astrophysical Masers: Unlocking the Mysteries of the Universe}
\editors{A. Tarchi, M.J. Reid, \& P. Castangia, eds.}
\begin{document}

\maketitle

\begin{abstract}

The Submillimeter Array (SMA) has been used to image the emission from radio recombination lines of hydrogen at subarcsecond angular resolution from the young high-mass star MWC349A in the H26$\alpha$, H30$\alpha$, and H31$\alpha$ transitions at 353, 232, and 211~GHz, respectively. Emission was seen over a range of 80\kms\ in velocity and 50~mas (corresponding to 60~AU for a distance of 1200~pc). The emission at each frequency has two distinct components, one from gas in a nearly edge-on annular disk structure in Keplerian motion, and another from gas lifted off the disk at distances of up to about 25~AU from the star. The slopes of the position-velocity (PV) curves for the disk emission show a monotonic progression of the emission radius with frequency with relative radii of $0.85\pm0.04$, 1, and $1.02\pm0.01$ for the H26$\alpha$, H30$\alpha$, and H31$\alpha$ transitions, respectively. This trend is consistent with theoretical excitation models of maser emission from a region where the density decreases with radius and the lower transitions are preferentially excited at higher densities. The mass is difficult to estimate from the PV diagrams because the wind components dominate the emission at the disk edges. The mass estimate is constrained to be only in the range of 10--30 solar masses. The distribution of the wind emission among the transitions is surprisingly different, which reflects its sensitivity to excitation conditions. The wind probably extracts significant angular momentum from the system.

\keywords{masers -- stars: emission-line, Be -- stars: winds, outflows -- radio lines: stars}
\end{abstract}

\firstsection 
\section{Introduction}

MWC349A is a highly unusual object that has intrigued and confused astronomers since its identification as a peculiar emission line star in the Mount Wilson Catalogue of Be stars by \cite[Merrill \& Burwell (1933)]{MerrillBurwell33}. The evolutionary state of the star, its relation to its apparent companion, MWC349B, and the nature of its circumstellar neutral and ionized envelopes are subjects of lively debate. For a comprehensive description of its characteristics, see, for example, \cite[Gvaramadze \& Menten (2012)]{GvaramadzeMenten12}. Its distance is usually taken to be 1.2~kpc (\cite[Cohen \etal\ 1985]{Cohenetal85}), which is derived from the spectral type (B0III) of MWC349B. On the other hand, its distance could be as high as 1.7~kpc, if it is within the Cygnus OB2 association. Its bolometric luminosity based on the nearer distance is about $40,000~L_\odot$, which corresponds to a ZAMS mass of about $40~M_\odot$. MWC349A is among the brightest radio stars known because of emission from a biconical region of ionized gas, which has a nearly perfect $\nu^{0.6}$ power law spectrum from below 1~GHz to above 1~THz. This appears to be a classic case of a constant velocity ionized outflow in which the electron density decreases as the square of the distance from the source.
\begin{figure}[th]
\begin{center}
 \includegraphics[trim=0cm 4.cm 4cm 3.0cm,  clip=true, width=5.3in]{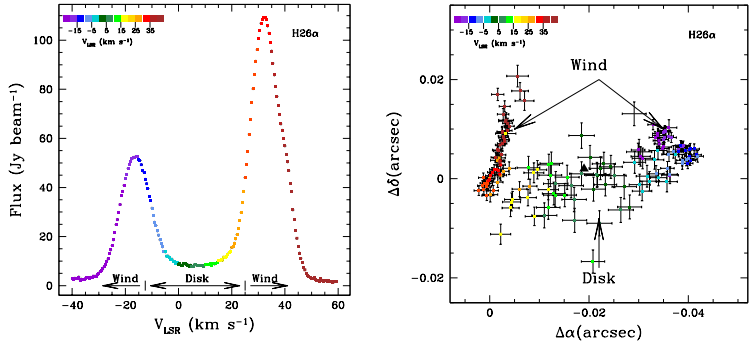} 
\vspace{-3.3cm}
 \caption{Observations of the H26$\alpha$ RRL in MWC349A made on Oct.~12, 2015, with the SMA. (left) Spectrum with 0.35\kms\ resolution. The velocity ranges of the origin of the dominant emission are shown. (right) The positions of the emission as a function of velocity determined by centroid analysis. The stellar position is shown as a black triangle. At a distance of 1200~pc, $1''=1200$~AU. Adapted from \cite[Zhang \etal\ (2017)]{Zhangetal17}.}
   \label{fig1}
\end{center}
\end{figure}

Of particular interest to this conference is that MWC349A hosts one of only three known examples of radio recombination line (RRL) maser emission (\cite[Mart{\'{\i}}n-Pintado \etal\ 1989]{MPintadoetal89}). At large principal quantum numbers, e.g., $n>41$, the spectral profile is a simple Gaussian function, whereas at smaller values of $n$, down to at least 15, the profile develops a prominent double structure (see Fig.~1) suggestive of a rotating disk viewed edge-on with limb brightening. It is interesting that RRL masers are apparently so rare, since the structure of the hydrogen energy levels seem particularly suited to generate maser action. In the 1930s, Menzel (e.g., \cite[Menzel 1937]{Menzel37}) made extensive theoretical calculations of the expected departures of the hydrogen level populations from their expected LTE values. Menzel, in the 1937 paper, noticed that the effects he calculated could lead to population inversion but said, ``The condition may conceivably arise when the [excitation temperature] turns out to be negative. ... The process merely puts energy back into the original beam, as if the atmosphere had a negative opacity. This extreme will probably never occur in practice." Menzel's calculations became important in interpreting the RRLs from HII regions starting in the mid-1960s because the kinetic temperatures inferred from the line-to-continuum ratios gave kinetic temperatures that were much lower than measured by other means (\cite[Dupree \& Goldberg 1970]{DupreeGoldberg70}). In these cases, the non-LTE effects were significant but did not produce outright population inversion.

The proclivity of the hydrogen atom to natural maser action can be understood from a simple model in which hydrogen atoms in the envelope of a massive star are ionized by the strong ultraviolet radiation field and then recombine and cascade down the principal energy levels to the ground state. The Einstein A coefficient, which gives the rate of spontaneous emissions, varies approximately as $n^{-5}$. If collisions are negligible, then the population levels will be largely controlled by the spontaneous emissions. Hence, for a given level, the transition rate into it from the level above is slightly less than the transition rate to the next lower level. This ``bottleneck" effect means that all levels will be inverted. Collisions disrupt this inversion, and higher densities are required to quench the inversion as $n$ decreases (see, e.g., \cite[Strelnitski \etal\ 1996]{Strelnitskietal96}). This situation is the opposite of that found in molecules undergoing rotational level transitions for which $A\sim J^3$, where $J$ is the rotational quantum number. Of course, detectable maser emission requires high negative optical depth in addition to population inversion.

The critical observational development in the study of the RRL line in MWC349A came with the measurement of the spatial separation of the two most prominent components of the spectrum with the OVRO interferometer in the H30$\alpha$ line by \cite[Planesas \etal\ (1992)]{Planesasetal92}, which gave a separation of 65~mas for the components separated by 48\kms. Interpreting this in a Keplerian disk model gave a mass $M=v^2R/G$ of $25\pm6~M_\odot$. This work was followed by SMA observations of the detailed distribution of masers, which showed the features following a disk structure (\cite[Weintroub \etal\ 2008]{Weintroubetal08}; \cite[Zhang \etal\ 2017]{Zhangetal17}). \cite[Mart{\'{\i}}n-Pintado \etal\ (2011)]{MPintadoetal11} used the PdBI to show that the velocity components greater than 15\kms\ from the systemic velocity were positioned off the disk and probably attributable to gas in a wind created by photoionization of the disk.

\begin{figure}[t]
\begin{center}
\includegraphics[width=4in]{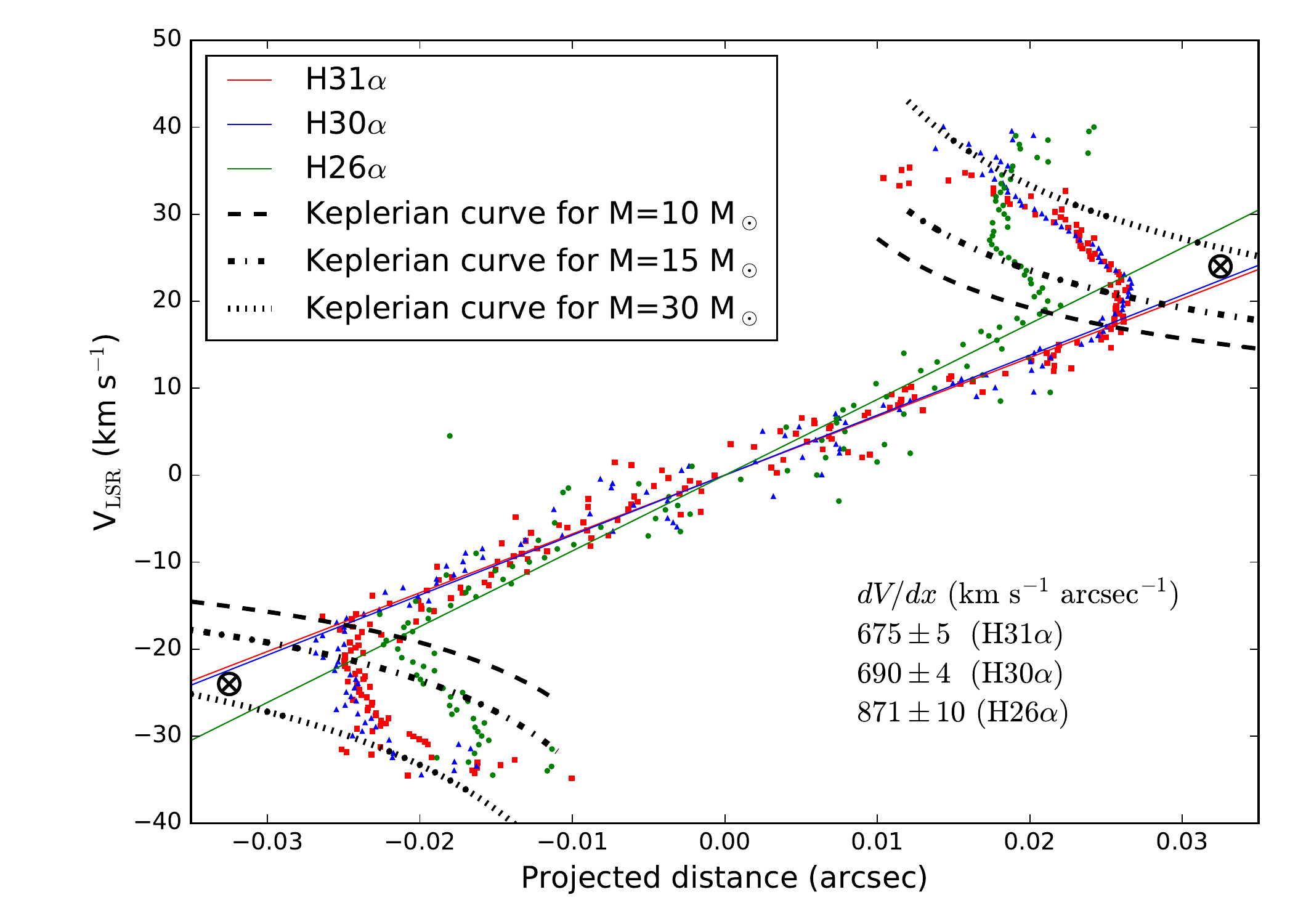} 
 \caption{The PV diagram for the maser emission centroids for MWC349A for the H26$\alpha$ (green), H30$\alpha$ (blue), and H31$\alpha$ (red) transitions. The increasing slope of the PV curves with decreasing $n$ is expected for a masing disk with decreasing density with increasing radius. The Keplerian curves for masses of 10, 15, and $30~M_\odot$ are shown as dashed lines. The two-point measurement of \cite[Planesas \etal\ (1992)]{Planesasetal92} is shown by circled crosses. Adapted from \cite[Emery (2017)]{Emery17}.}
   \label{fig2}
\end{center}
\end{figure}

\begin{figure}[t]
\begin{center}
 \includegraphics[width=3in]{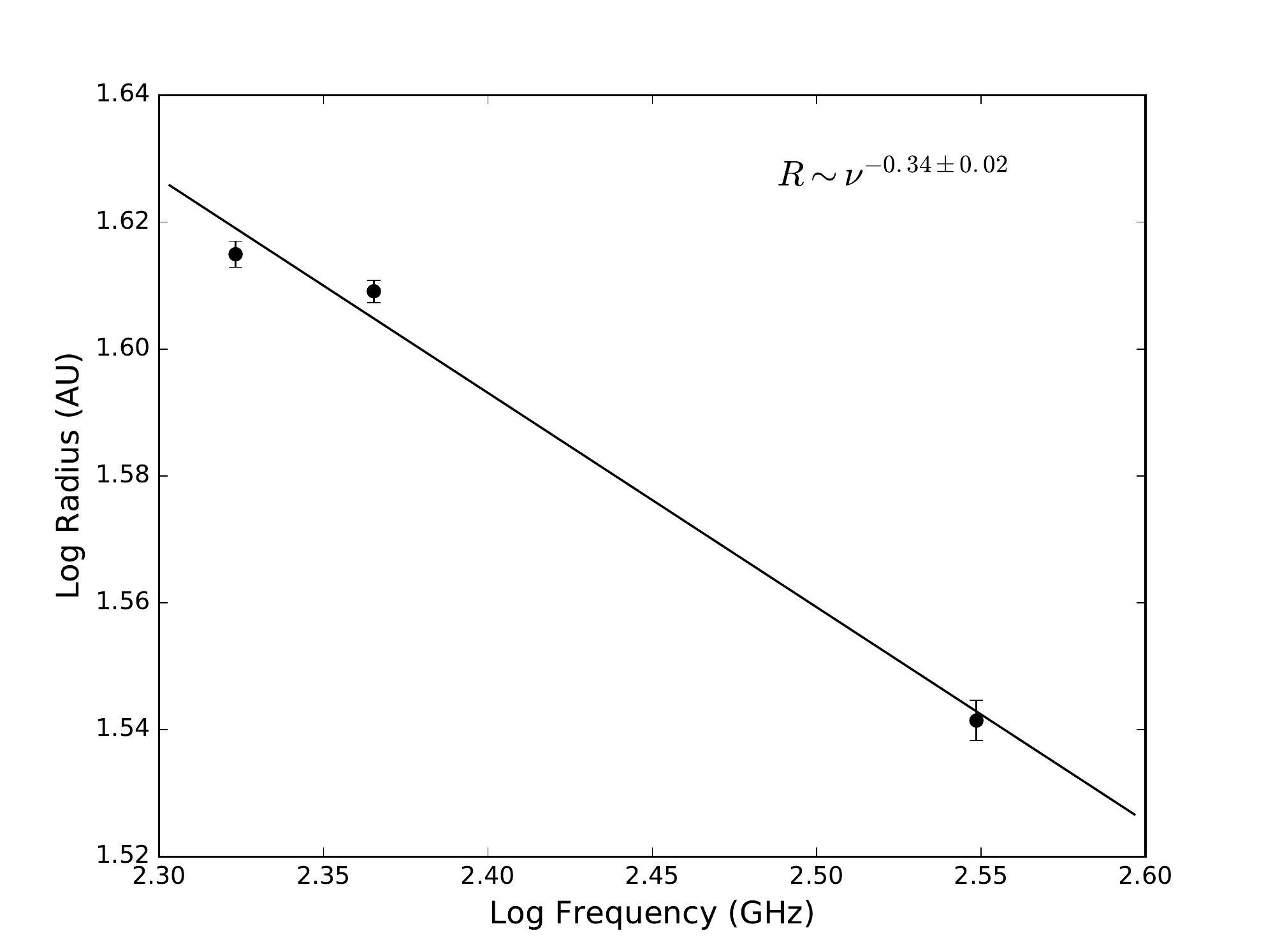} 
 \caption{The annular radius of each RRL maser as a function of transition frequency derived from the slope of the PV diagram. A power-law fit is also plotted. From \cite[Emery (2017)]{Emery17}.}
   \label{fig3}
\end{center}
\end{figure}

We report here on observations of the H26$\alpha$, H30$\alpha$, and H31$\alpha$ lines. Figure~\ref{fig1} shows the spectra and spatial distribution for the H26$\alpha$ transition. The position of the emission at each velocity was determined by a centroiding analysis, which is strictly appropriate only for an unresolved distribution. The formal position accuracy can exceed the resolution of the interferometer by a factor of about the signal-to-noise ratio. If multiple components are present, the centroiding analysis gives some sort of flux-weighted mean position whose interpretation can be unclear and misleading. The elongated distribution corresponding to features within 15\kms\ for the systemic velocity corresponds to a nearly edge-on disk. The position-velocity (PV) diagrams for emission in all three transitions are shown in Figure~\ref{fig2}. The emission is expected to peak in an annular ring, where the density is optimal for maser action. The radii can be estimated from the slope of the PV plot since $V=\sqrt{GM/R}\sin \phi=\sqrt{GM/R^3}x$, where $G$~is the gravitational constant, $R$~is the radius, $\phi$~is the azimuthal angle in the disk measured from the line-of-sight, and $x$~is the projected distance. The ratios of the radii can be determined with respect to the radius of the H30$\alpha$ emission without knowledge of the stellar mass. The relative radii of the H26$\alpha$, H30$\alpha$, and H31$\alpha$ emission peaks are $0.85\pm0.04$, 1 and $1.02\pm0.01$, respectively, and for a mass of $25~M_\odot$, the radii are 35, 40, and 41~AU. These results are shown in Figure~\ref{fig3}. The radii scale as $\nu^{-0.34\pm0.02}$ or $n^{1.0\pm0.06}$. This dependence is in reasonable agreement with theoretical models (e.g., \cite[Strelnitski \etal\ 1996]{Strelnitskietal96}; \cite[Thum \etal\ 1994]{Thumetal94}).

The mass is rather difficult to estimate from the data. The two-point Keplerian-model analysis (\cite[Planesas \etal\ 1992]{Planesasetal92}) gives $25\pm6~M_\odot$ (see Fig.~\ref{fig2}). If the emission in the straight line portion of the PV diagrams shown in Figure~\ref{fig2} is assumed to trace the whole range of disk azimuth angles ($\pm90^\circ$), e.g., limb-to-limb, then the mass estimate would be only $10~M_\odot$. However, a more reasonable interpretation of the data is that a centroiding process obscures the disk emission at $|\phi|>50^\circ$ because of the dominance of the wind emission. In this case, the true mass could be easily as large as $30~M_\odot$.

Finally, we note that the wind emission seems to be excited along rather different trajectories in three transitions. This is probably due to a particular sensitivity of the maser emission in any transition to density distribution. The origin of this wind is discussed by \cite[B{\'a}ez-Rubio \etal\ (2014)]{BaezRubioetal14}.

\smallskip

We thank Vladimir Strelnitski for pedagogical discussions about the population distribution in the hydrogen atom and Alejandro B{\'a}ez-Rubio for helpful discussions about radiative transfer modeling.

\vspace{-0.4cm}

\def\apj       {\textit{ApJ}}
\def\apjl      {\textit{ApJ} (Letters)}
\def\aap      {\textit{A\&A}}
\def\araa     {\textit{ARAA}}

\end{document}